\newcommand{\beq}{\begin{equation}}
\newcommand{\eeq}{\end{equation}}
\newcommand{\bea}{\begin{eqnarray}}
\newcommand{\eea}{\end{eqnarray}}

\newcommand{\gsim}{\lower.7ex\hbox{$\;\stackrel{\textstyle>}{\sim}\;$}}
\newcommand{\lsim}{\lower.7ex\hbox{$\;\stackrel{\textstyle<}{\sim}\;$}}

\documentclass[aps,prev,preprintnumbers,floatfix,nofootinbib]{revtex4-1}
\usepackage{graphicx}
\usepackage{epstopdf}
\usepackage{mathrsfs}
\usepackage{amssymb}
\usepackage{verbatim}
\usepackage{color}
\usepackage{multirow}
\usepackage{amsmath}
\usepackage{float}
\usepackage[normalem]{ulem}
\usepackage[lofdepth,lotdepth,caption=false]{subfig}

\def\stacksymbols #1#2#3#4{\def\theguybelow{#2}
    \def\vp{\lower#3pt}
    \def\sp{\baselineskip0pt\lineskip#4pt}
    \mathrel{\mathpalette\intermediary#1}}

\def\intermediary#1#2{\vp\vbox{\sp
     \everycr={}\tabskip0pt
     \halign{$\mathsurround0pt#1\hfil##\hfil$\crcr#2\crcr
              \theguybelow\crcr}}}

\def\be{\begin{equation}}
\def\ee{\end{equation}}
\def\bea{\begin{eqnarray}}
\def\eea{\end{eqnarray}}
\begin{document}

\vspace*{1mm}

\title{Unravelling the richness of dark sector by FASER$\nu$}
\author{Pouya Bakhti$^a$}
\email{pouya$_$bakhti@ipm.ir}
\author{Yasaman Farzan$^{a,b}$}
\email{yasaman@theory.ipm.ac.ir}
\author{Silvia Pascoli$^c$}

\vspace{0.2cm}

\affiliation{
${}^a$School of physics, Institute for Research in Fundamental Sciences (IPM),\\
P.O. Box 19395-5531, Tehran, Iran\\
${}^b$The Abdus Salam ICTP, Strada Costiera 11, 34151, Trieste, Italy\\
${}^c$ 
Institute for Particle Physics Phenomenology, Department of Physics, Durham University, Durham
DH1 3LE, U.K.}

\begin{abstract} 
FASER$\nu$ is a newly proposed experiment which will take data in  run III of the LHC during 2021-2023. It will be located in front of the FASER detector, 480~m away from the ATLAS interaction point  in the forward direction. Its main goal is to detect neutrinos of all flavors produced at the interaction point with superb precision in reconstructing charged tracks. This capability makes FASER$\nu$ an ideal setup for uncovering the pattern and properties of a light dark sector. We demonstrate this capability for a well-motivated class of  models with a dark matter candidate of mass around a few GeV. Dark matter annihilates to a pair of intermediate neutral particles that subsequently decay into the standard model charged fermions.
We show how FASER$\nu$ can shed light on the structure of the dark sector by unravelling the decay chain within such models.
\end{abstract}

\maketitle
\section{Introduction}

In recent years, inspired by the richness of the standard model particle content  and motivated by certain observations on structure formation, there has been a paradigm shift in the dark matter model building from minimality to having an extended dark sector with new interactions among dark particles.  Given the null results from the direct and indirect  dark matter search experiments, there is a growing interest in dark sectors at the GeV scale which interact too weakly to have been discovered so far.  Low energy experiments with high luminosity are the best setups to search for such feebly interacting light particles. The run III of the LHC experiment will enjoy very high luminosity ($150~{fb}^{-1}$) of protons colliding at a center of mass energy of $\sqrt{s}={14}$ TeV, followed by a further luminosity increase of $O(10)$. Protons, being composite particles, abundantly contain partons which carry only a small fraction ($x$) of the proton momentum. If two  partons with $x_1 \sim 0.1$ and $x_2 \sim {\rm GeV}^2/(s x_1)$ collide, they can produce new light particles with a mass around GeV highly collimated in the forward direction.
As a result, detectors such as FASER \cite{Ariga:2018pin} or FASER$\nu$ \cite{Abreu:2019yak}, which are going to be installed in the forward direction at a distance of 480 m from the Interaction Point (IP) of ATLAS, can be considered low energy high luminosity experiments ideal for this type of searches. 

The potential of the FASER experiment has been widely discussed in the literature \cite{Ariga:2018uku, Alimena:2019zri}.
Initial studies focussed on light scalar singlet mixed with the SM Higgs \cite{Feng:2017vli}, subsequently extended to a connection to the inflaton \cite{Okada:2019opp} and to an improved sensitivity to the Higgs portal \cite{Boiarska:2019vid}. The possibility to search for heavy neutral leptons has been considered, both when they arise from heavy meson decays \cite{Kling:2018wct,Helo:2018qej} and  in the presence of new $B-L$  gauge interactions \cite{Deppisch:2019kvs}. FASER sensitivity to dark matter particles and axions/dark photons, and their possible connections, has been explored within various scenarios \cite{Feng:2018noy,Mohapatra:2019ysk,Berlin:2018jbm}. Complex dark sectors involving multiple light particles can also lead to unique signatures in FASER  \cite{Jodlowski:2019ycu}. Finally, the imprint of $R$-parity violating supersymmetric models in which unstable neutralinos are produced in heavy meson decays  has been looked at in Ref. \cite{Dercks:2018eua}. FASER$\nu$ is a more recent proposal \cite{Abreu:2019yak} and its unique capabilities have not  been fully explored, yet (see, however, \cite{Pouya,Kling:2020iar}). The main purpose of FASER$\nu$ is to detect neutrinos produced at the interaction point.
To reconstruct the $\nu_\tau$
charged current interactions FASER$\nu$ is designed to have superb spatial
as well as position resolution of respectively 50 nm and 0.4 $\mu m$ which makes it capable of reconstructing the direction of the tracks of the charged particles with a remarkable precision.

If the mass of dark matter is smaller than 10 GeV and its main coupling is of the electroweak type, the cross-section of the annihilation of dark matter pair into SM particles can be estimated as $\sigma \propto G_F^2 m_{DM}^2 \ll pb$ so as shown in 1977 by Lee and Weinberg \cite{Lee:1977ua}, the standard freeze-out scenario cannot yield the observed abundance for the dark matter. Thus, light dark matter  needs another interaction mode. One possibility is proposed within the SLIM scenario \cite{SLIM}. Another attractive scenario is the annihilation of dark matter pair to a pair of intermediate particles which are singlets of $SU(2)\times U(1)$ and subsequently decay into SM particles.
For such Dark Matter (DM) scenarios, FASER$\nu$ will enjoy  spectacular signals. The aim of the present paper is to demonstrate how the superb capabilities of FASER$\nu$ in the track and vertex reconstruction can help to learn about the intricate structure of the decay chain of the dark sector. For concreteness, we develop a class of dark matter models that go through decay chain.
We then show how the scenario fits in the freeze-out paradigm for the DM production in the early universe and provides a viable dark matter model for  both cases that the DM particle is a fermion or a scalar.

 In  models of our interest,  intermediate particles $X'$ are produced at the Interaction Point (IP) and subsequently chain-decay leading to unique signatures at FASER$\nu$.
After presenting different  scenarios for  $X'$, we focus on the specific case of a pseudoscalar $X'$ with an axion type interaction with the gluons of form
\begin{equation} \label{coupling}
\frac{X' G_{\mu \nu}^i G_{\alpha \beta}^i \epsilon^{\mu \nu \alpha \beta}}{\Lambda},
\end{equation} where $G_{\alpha \beta}^i$ is the gluon field strength. This kind of interaction has been introduced in axion models motivated by the solutions for the QCD CP-problem. 
Typically, the mass of the axion is suppressed by the inverse of the coupling $\Lambda$. The stringent bounds on the axion scale then imply the axion mass to be very light. In recent years, increasing interest has arisen on models which allow for heavier masses, in particular above the MeV scale so that astrophysical bounds become much weaker \cite{others}.
Recently, Ref. \cite{Hook:2019qoh} extends the minimal axion picture to ameliorate the so-called ``axion quality problem", that is the sensitivity of the axion mechanism to higher dimensional corrections from heavy states. In this model, the mass of $X'$ is not suppressed by $1/\Lambda$ and can be in the range 0.1 GeV to 10 TeV. In this paper, we focus on $X'$ with a mass in the range of few GeV-few 10 GeV. At the interaction point, such $X'$ can be produced abundantly in the forward direction. We turn on a coupling between $X'$ and the dark sector such that $X'$ immediately decays into new invisible particles with dominant branching ratio. We then study possible signatures of the dark sector in FASER$\nu$.

 Within our scenario, the $X'$ particles immediately decay to the neutral $X\bar{X}$ pair which are singlets of the standard model and can penetrate the rock and concrete between the IP and FASER$\nu$. The $X$ particles have a relatively long lifetime enough for them to reach the FASER$\nu$ detector. They eventually decay into dark matter $Y$ and a pair of neutral $\eta$ and $\bar{\eta}$ particles.\footnote{As we shall see in sec \ref{ConnectionTODM}, for fermionic dark matter ($Y$), $\eta$ has to be a complex field in order to have a successful freeze-out scenario for the dark matter production in the early universe but for  scalar a $Y$, $\eta$ can be either complex or real. For real $\eta$, $\bar{\eta}$ will of course be the same as $\eta$.} $Y$ leaves the detector as missing energy but $\eta$ and $\bar{\eta}$ decay into pairs of charged leptons  which can be detected.  The signal will therefore be two pairs of charged leptons plus missing transverse momentum associated with the dark matter production. We show that by reconstructing the invariant mass of the charged lepton pairs, we can veto the neutrino beam induced background and extract information on the mass of $\eta$. We find that for 
$10^{-10}~ {\rm sec}<\tau_X<10^{-4}~ {\rm sec}$, yet-unexplored values of $\Lambda$ can be tested by FASER$\nu$. We discuss how other parameters of the model such as the $\eta$ lifetime can be extracted, thanks to the unique capabilities of FASER$\nu$.

In sect. \ref{three-scenarios}, we describe three different possibilities for the $X'$ particle. We then discuss the lifetime of $X'$. In sect. \ref{production}, we compute the production cross-section of $X'$  through the coupling in Eq. (\ref{coupling}) and then derive the energy spectrum of the $X$ particles from the $X'$ decay.  In sect. \ref{dark-sector}, we show how FASER$\nu$ and FASER can constrain the coupling of $X'$ to the gluons ($\Lambda$)  when $X'$ decays dominantly to the singlets $X$ and $\bar{X}$ rather than to the gluons. In sect. \ref{chain}, we proceed with different scenarios for the $X$ decays with particular attention to the dark matter. We describe the signatures of each scenario and demonstrate how the unique capabilities of FASER$\nu$ can be sensitive to them. We then discuss extracting the parameters of the model such as the mass of the intermediate particles and their  lifetimes. In sect. \ref{ConnectionTODM}, we build a light dark matter model  based on the thermal freeze-out dark matter production mechanism that embed the introduced particles. In sect. \ref{summary}, we summarize our results and briefly discuss how other future detectors can test the model.



\section{Different scenarios for the intermediate particle, $X'$ \label{three-scenarios}}

In this section, we briefly present three scenarios for the intermediate particles, $X'$. In order to have an observable signal at FASER$\nu$, we require $X'$ to couple to the SM not too weakly so that it can be copiously produced at the IP.

\begin{itemize}
\item $X'$ is a gauge boson of mass of few 100~MeV- few 10 GeV associated with the anomaly free symmetry $B-a_e L_e-a_\mu L_\mu -a_\tau L_\tau$ (with $a_e+a_\mu+a_\tau=3$) or a dark photon mixed with ordinary photons. Such possibilities are studied in
\cite{Deppisch:2019kvs, Mohapatra:2019ysk, Berlin:2018jbm, Jodlowski:2019ycu}.
\item $X'$ is a scalar coupled to the quarks. Refs. \cite{Feng:2018noy, Mohapatra:2019ysk, Berlin:2018jbm}
consider a singlet scalar mixed with the SM Higgs.
As a result, the effective coupling of this scalar to
the heavier generations will be stronger and their main production mode at the LHC would be the decay of heavy mesons containing the $c$ or $b$ quarks. Because of the smallness of the coupling to the first generation,
the scalar in the scenario cannot directly be produced via parton fusion at the IP. However, by introducing a new Higgs doublet whose Yukawa coupling to the first generation is large, one can obtain large coupling between the new light singlet scalar and the $u$ and $d$ quarks. To avoid large flavor changing effects, a symmetry should be invoked to guarantee that the coupling of the new Higgs to the quarks is diagonal in the flavor space.
\item $X'$ is a scalar which couples to the gluons via the interaction shown in Eq.~(\ref{coupling}). This is the case that we are going to explore in this paper.\\
	The  rate of the decay of $X'$ into a gluon pair is given by \cite{Hook:2019qoh}
	\begin{equation} \Gamma (X' \to gg)=3\times 10^{11} ~{\rm sec}^{-1} \left(\frac{2\times 10^4~{\rm GeV}}{\Lambda}\right)^2 \left( \frac{m_{X'}}{3~{\rm GeV}}\right)^3.\end{equation}
	$X'$ should also couple to the singlet  $X$ particles. We have not specified whether $X$ is scalar or fermion. For scalar $X$, we may introduce trilinear coupling of type $\lambda m_{X'} X' \bar{X}X$ and for fermionic $X$, we may write a Yukawa coupling of form $\lambda X'\bar{X}X$.
	In order for $X'$ to decay into a $X\bar{X}$ pair faster than into $gg$, the $\lambda$ coupling  should be larger than $O(10^{-6})$. $X'$,  with an energy of 100 GeV- few TeV, will decay after propagating less than 1~m so it will decay before reaching the detectors. At one loop level, the $\lambda$ coupling induces a contribution to $ m_{X'}^2$ less than  $O(\lambda^2 m_{X'}^2/16 \pi^2)$ which is negligible. However, the same coupling will also induce a tadpole contribution for $X'$ at one loop level proportional to $\lambda$ which can be problematic for fixing the vacuum expectation value of  $ X'$ to zero and solving the QCD $\theta$-term. Since the tadpole contribution is linear in $\lambda$, we may introduce a replica of  the $X$ particle with the same mass but opposite coupling to $X'$ to cancel the radiative tadpole. Addition of such replica will not alter the signatures at FASER$\nu$ so we will not discuss it  any further.
We just note in passing that  if the replica of $X$ also decays within the detector with a decay mode similar to that of $X$ (as the symmerty between $X$ and its replica suggests), the phenomenology that we are discussing will be exactly the same. If the replica decays before reaching FASER$\nu$ or if it is stable, the only difference in the prediction for FASER$\nu$ will be the reduction of the statistics by half. In other words, we would recover the same results by replacing $\Lambda \to \sqrt{2} \Lambda$.
\end{itemize}

\section{Production of $X'$ \label{production} }

The $X'$ particles coupled to the gluons via the interaction in Eq. (\ref{coupling}) can be produced in the gluon fusion at the Interaction Point (IP): $g(p_1)+ g(p_2)\to X'(p_1+p_2)$. Neglecting the transverse momentum of the gluons, we may write
\be p_1=p(x_1,0,0,x_1) \ \ \ \ {\rm and} \ \ \ \ p_2=p(x_2,0,0,-x_2) \label{momenta} \ee
where $p\simeq 7$ TeV is the momentum of the proton. Energy-momentum conservation then implies
$$ m_{X'}=p \sqrt{4 x_1x_2} \ \ \ \ \ {\rm where} \ \ \ \ \ \frac{m_{X'}^2}{4p^2}
<x_1,x_2<1.$$
The square of the scattering amplitude after
averaging over the spins and the colors of the initial gluons is given by \footnote{Within the axion models, the coupling is parameterized as $\Lambda =8 \pi f_a/\alpha_{QCD}$ where $\alpha_{QCD}$ is the QCD structure constant. In general, $\alpha_{QCD}$ runs with the energy scale. However, considering the fact that for $ g+g \to X'$, the energy scale is fixed by $m_{X'}$ we do not need to worry about the running of the coupling and we may simply describe it by a constant $\Lambda$.}
$$\langle |{M}^2|\rangle =\frac{2 p^4 x_1^2 x_2^2}{\Lambda^2}. $$
The cross section of two gluons with energies $E_1=px_1$ and $E_2=px_2$ fusing into $X'$ is the following
\be \sigma(x_1,x_2) =2\pi\int \frac{\langle |{M}^2|\rangle}{2E_12E_2|v_2-v_1|} \delta^4(p_1+p_2-p_X')\frac{d^3P_{X'}}{2E_{X'}} =\frac{\pi}{4 \Lambda^2}\frac{x_1x_2}{x_1+x_2}\delta[x_1+x_2-\sqrt{(x_2-x_1)^2+m_{X'}^2/p^2}].
\label{sigma}\ee
Convoluting with the gluon parton distribution function, $F_g$, we find
\be \frac{d \sigma_{tot}(E_{X'})}{d E_{X'}}=\frac{\pi}{32 \Lambda^2}\frac{m_{X'}^2}{p^2}\frac{1}{P_{X'}} F_g(\frac{E_{X'}+P_{X'}}{2p})
F_g(\frac{E_{X'}-P_{X'}}{2p}),\ee
in which $P_{X'}=\sqrt{E_{X'}^2-m_{X'}^2}$.
Neglecting the transverse momenta
implies that all 
the $X'$ particles will be emitted parallel to the beamline.  The partons inside a proton should have a transverse momentum comparable to the inverse of the proton size: $p_t=200$ MeV. Taking $m_{X'}>$GeV, this momentum can be neglected in the computation of the cross-section but it will lead to the emission of  $X'$ within a small solid angle of $\pi\theta_t^2$ where $\theta_t=p_t/E_{X'}$.
The typical momentum of the protons within a bunch is smaller than MeV and can be neglected.

Let us denote the angle subtending the detector from the point at which $X'$ decays into $X\bar{X}$ with $\theta_d$: $\tan \theta_d=(0.25~{\rm m})/d$
where $d$
is the distance between the point at which $X'$ decays and the detector.  If the $X'$ lifetime is shorter than $10^{-11}$ sec, the distance traveled by $X'$ will be shorter than few meters so it can be neglected in comparison to the distance between the IP and the detector.
Thus, for a prompt decay of $X'$, $d=480$ m and therefore $\theta_d=5 \times 10^{-4}$.

Let us now compute the spectrum of the $X$ particles which are produced via $X' \to X \bar{X}$. In the rest frame of $X'$, the momentum of the $X$ particle will be $$k=(m_{X'}^2/4-m_X^2)^{1/2}.$$ The velocity of $X'$ in the lab frame is $v_{X'}=(1-m_{X'}^2/E_{X'}^2)^{1/2}$.

As long as $m_{X'}\sqrt{1-v_{X'}^2}/2<m_X$, we expect  $k<m_{X'}v_{X'}/2$ so the angle that the momentum of the daughter $X$ makes with the direction of the momentum of the mother $X'$ will be smaller than $$\theta_s=\arctan \left( \frac{2 k}{E_{X'}v_{X'} \sqrt{1-4k^2/(m_{X'}^2 v_{X'}^2)}} \right) .$$
For $p_t \ll k\sim m_{X'}/2$, this angle will however be larger than $\theta_t$ and the angular spread of the emitted $X$ can therefore be described by $\theta_s$.
For $m_X \to m_{X'}/2$, $k\ll m_X/2$ so there will be no angular dispersion and all the $X$ particles will be emitted in the direction of the $X'$ momentum.
To be more precise for
 $k\sim p_t\sim 200~{\rm MeV}\ll m_{X'}/2$, the produced $X$ particles will make an angle smaller than $2k/(\gamma m_{X'}v_{X'})\ll 1$ with the  $X'$ momentum so all the $X$ particles will practically be redistributed within $\pi \theta_t^2$ with
\be E_X=\frac{E_{X'}}{2}(1+\frac{2 k}{m_{X'}}) \label{ExEx'}. \ee
In the case of $k\gg p_t$, relation (\ref{ExEx'}) also holds valid 
for the fraction of the $X$ particles directed towards the detector, up to a correction of $O[p_t^2/(4 k m_{X'})]$.
In the both cases, the flux of the $X$ particles will be therefore given by
\be \label{spec}
\frac{d N(E_X)}{d E_X}= L f \frac{d\sigma_{tot}}{d E_{X'}}|_{E_X} \frac{2}{1+\sqrt{1-4m_X^2/m_{X'}^2}} \ee where $L$ is the integrated luminosity.
$f$ is the fraction of $X$ that reach the detector. For $p_t\sim  200~{\rm MeV}\sim k\ll m_{X'}/2$ and for $p_t\ll k\sim  m_{X'}/2$, we can respectively write $f=\theta_d^2/\theta_t^2$ and  $f=\theta_d^2/(2 (1-\cos \theta_s))$.

{ $X'$ can also be produced in the $2 \to 2$ processes such as $g+g \to X' +g$ or $g+q \to X' +q$.  In fact, for the high $p_T$ processes that emit $X'$ in the direction of the ATLAS detector, these processes will dominate the $X'$ production \cite{Hook:2019qoh}. The $X'$ that are produced by the interaction of partons with $s=2\sqrt{x_1 x_2}p \gg m_{X'}$  via the $2 \to n$ processes with $n>1$   will be emitted in a wide solid angle. Practically only partons for which $2 \sqrt{x_1x_2} p 
	\sim m_{X'}$ produce $X'$ in the direction reaching FASER$\nu$. Taking the cross section of $2 \to 2$ process of order of $\alpha_{QCD}/\Lambda^2$, we find that $d N(E_X)/dE_X|_{2\to 2}$ cannot be larger than 20\% of $d N(E_X)/dE_X|_{2\to 1}$ in Eq. (\ref{spec}).  We therefore neglect the $2 \to 2$ processes in our analysis but in a more elaborate analysis, these processes should be also taken into account. They will increase the $X'$ production. As a result, our simplified computation gives a more conservative bound than the full production computation.}

\begin{figure}[h]
\hspace{0cm}
\includegraphics[width=0.8\textwidth, height=0.6\textwidth]{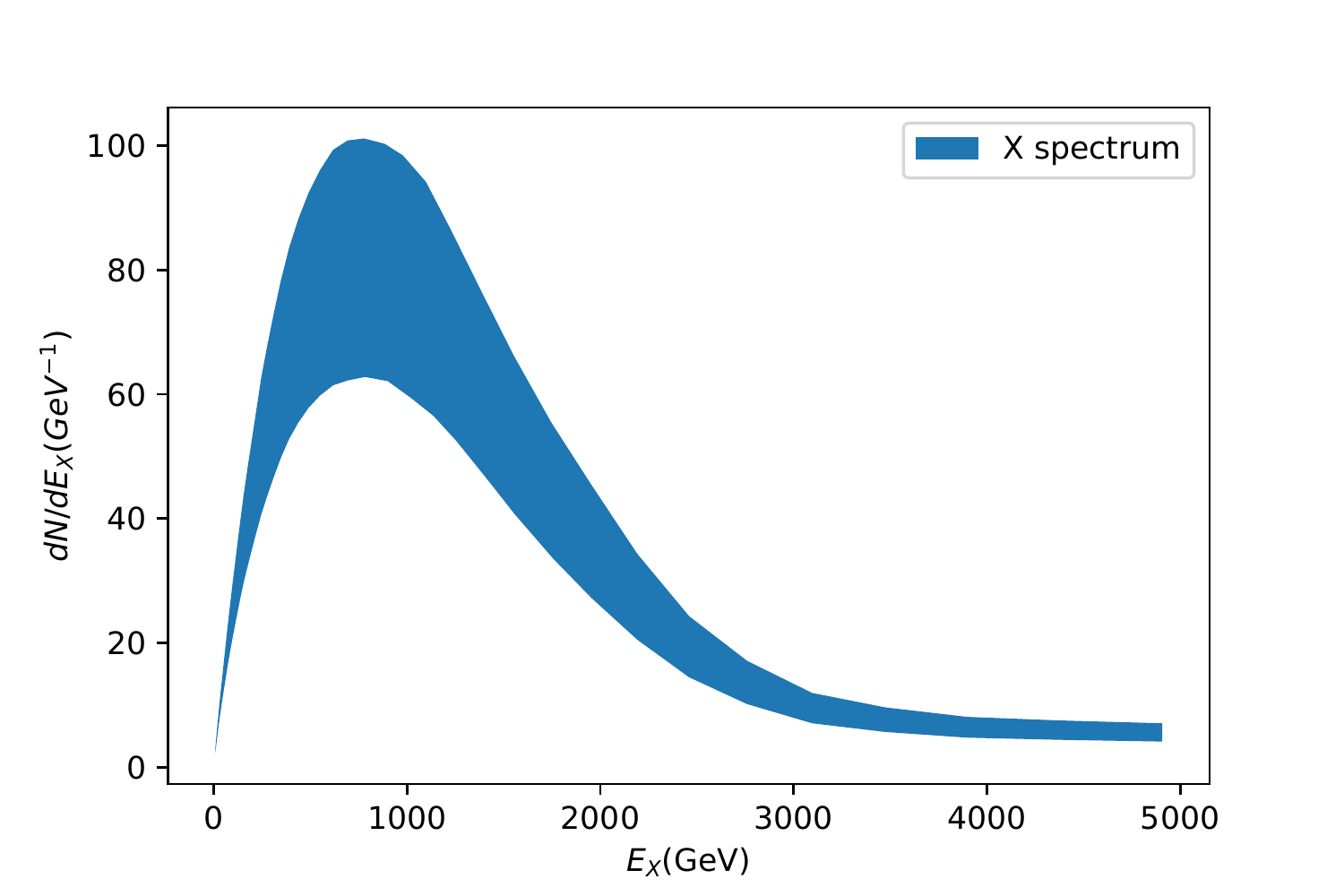}
\caption[...]{
The spectrum of $X$ particle emitted towards the FASER$\nu$ detector during the run III of LHC with an integrated luminosity of 150 fb$^{-1}$ and a center of mass energy for colliding protons of $\sqrt{s}={14}$ TeV. We have taken $m_{X'}=3$ GeV, $m_X=1$ GeV and $\Lambda=2 \times 10^4$ GeV. The gluon distribution function is taken from \cite{Rojo:2015acz}. The band reflects the uncertainties on the gluon distribution function especially at low momentum fractions. 
\label{spectrum}
}
\end{figure}

Fig.~\ref{spectrum} shows the spectrum of $X$ directed towards the detector during the run III of the LHC with an integrated luminosity of 150 fb$^{-1}$ and a center of mass energy for colliding protons of $\sqrt{s}={14}$ TeV. We have taken typical values $m_{X'}=3$ GeV, $m_X=1$ GeV and therefore
$k=1.12$ GeV. For the coupling, we have taken $\Lambda= 2\times 10^4$ GeV which is close to the minimal allowed values for $m_{X'}\sim 3 $ GeV \cite{Hook:2019qoh} . We have taken the parton distribution function from \cite{Buckley:2014ana,Rojo:2015acz} at $Q^2 =m_{X'}^2$.
The figure shows a peak around 1 TeV. To produce an $X$ particle with energy of $E_X=100~{\rm GeV}-1000~{\rm GeV}$, two gluons with momentum fractions of $x_1 = (E_{X'}/7~{\rm TeV})\sim [2 E_X/(1+2k/m_{X'})](1/7~{\rm TeV})\sim 0.1-0.01$ and $x_2= m_{X'}^2/[x_1 (14~{\rm TeV})^2]\sim 10^{-7}-10^{-6}$ should fuse together. At the momentum fractions of $x_2 \sim 10^{-7}-10^{-6}$, the gluon distribution function suffers from large uncertainties. The blue band in Fig. 1 reflects this uncertainty.
For the values of the parameters taken for Fig. 1, during the run III of the LHC $135000$ $X$ will be emitted in the direction of the detector with energies of few 100 GeV to few TeV. Of course, for larger $\Lambda$, the flux will be reduced by $O\left[ \left( \frac{2 \times 10^4~{\rm GeV}}{\Lambda} \right)^2 \right]$.

٪٪٪٪٪٪٪٪٪٪٪٪٪٪٪٪٪٪٪٪٪٪٪٪


\section{ Bounds on the $\Lambda$ coupling from FASER and FASER$\nu$\label{dark-sector}}
$X'$ decays into $X\bar{X}$ pair before traveling a sizable distance.
If the $X$ particles decay within FASER or FASER$\nu$ such that they produce a signal of dilepton,
the model can be probed. 
The dilepton  can be produced either directly by the $X$ decay or by the decay of the intermediate particles that are produced via the $X$ particles.\footnote{Instead of charged leptons, the intermediate particles can decay into quark pairs leading to dijets at the detector. Since the angular separation of the final charged fermion pair is going to be small, the dijets may not be separable. Hence, throughout the paper, we focus on dileptons in the final states to avoid such complications. }
{If the average distance that these intermediate particles travel before decay is between 1~mm and a few$\times 10$~cm, FASER cannot resolve the displaced vertex} but FASER$\nu$ will be able to do so. The main topic of the present paper is to focus on this particular capability of FASER$\nu$. Let us however first see what information FASER$\nu$ and FASER can give us on the $X$ lifetime. The number of events from the $X$ decay within the energy interval $E_X$ and $E_X+dE_X$ is given by 
\be \label{NomEvents} \frac{dN(E_X)}{dE_X}  \exp ^{-\frac{d }{\tau_X\gamma_X} }\left(\frac{s_z}{\tau_X \gamma_X}\right) dE_X\ee
in which $d=480$ m, $\tau_X$ is the lifetime of $X$ in its rest frame and $\gamma_X$ is the boost factor $\gamma_X=m_X/E_X$. The exponential gives the probability that $X$ reaches the detector. $s_z$ is the size of the detector along the beamline. For FASER$\nu$, $s_z=1.3$ m and for FASER2 (the upgrade of FASER for the high luminosity LHC), $s_z=5$ m. The ratio $s_z /(\tau_X \gamma_X)$ gives the probability of $X$ decaying inside the detector. Remembering that $dN/d E_X$ is proportional to $f\propto \theta_d^2 \propto $area of the detector,  Eq (\ref{NomEvents}) shows that the number of events will be proportional to the volume of the detector as expected.

\begin{figure}[h]
\hspace{0cm}
\includegraphics[width=0.8\textwidth, height=0.6\textwidth]{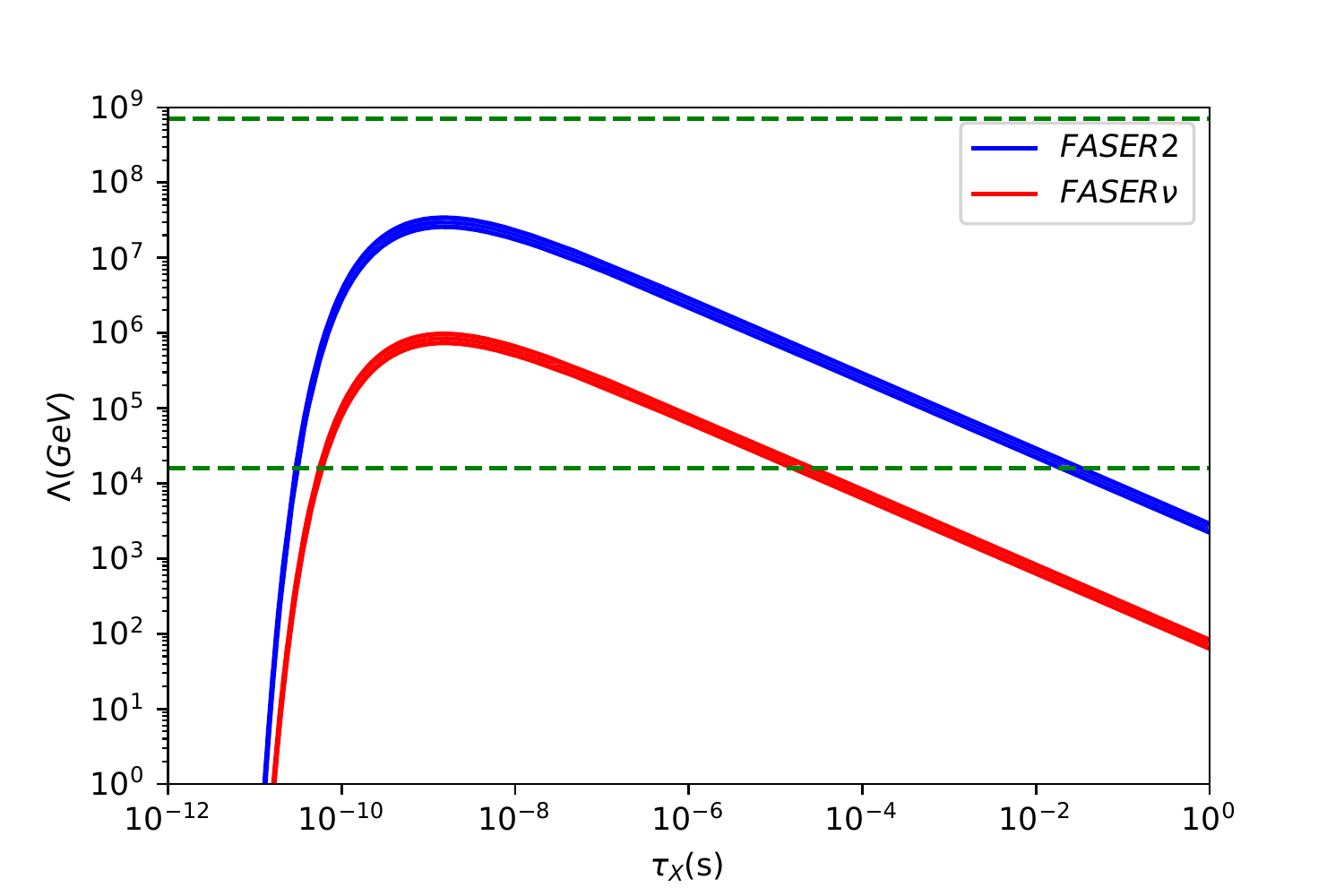}
\caption[...]{
Lower bound on $\Lambda$ versus the $X$ lifetime. The red band shows the bound by the FASER$\nu$ setup  in the run III of LHC   during 2021-2023 with a luminosity of 150 fb$^{-1}$ at a center of mass energy of colliding protons of $\sqrt{s}={14}$ TeV. The blue band shows the bound by the FASER2 setup  in the high luminosity run  of LHC   during 2026-2035 with a luminosity of 3000 fb$^{-1}$. We have taken $m_{X'}=3$ GeV and $m_X=1$ GeV. The gluon distribution function is taken from \cite{Rojo:2015acz}. The bands reflect the uncertainties on the gluon distribution function especially at low momentum fractions. The horizontal lines show the allowed range for $\Lambda$ \cite{Hook:2019qoh}.
\label{lower_bound}
}
\end{figure}

Taking $m_{X'}=3$ GeV and $m_X=1$ GeV, Fig. \ref{lower_bound} shows the lower bound that can be set on $\Lambda$ versus the $X$ lifetime in case no signal is observed. We have assumed zero background to the signal. We shall justify the assumption in the next section. We have assumed 100\% efficiency for the detection of the final SM fermions which according to \cite{Kodama:2007aa,Kobayashi:2012jb,Arrabito:2007rq} is a good assumption. The red band corresponds to the results from FASER$\nu$ with $L=150~{\rm fb}^{-1}$ taking into account the uncertainties in the gluon distribution function. The region below the red curve will be probed by run III of the LHC for our scenario where $X'$ decays
to a pair of $X$ and $\bar{X}$ which in turn lead to a background-free signal at FASER$\nu$. The shape of the curve can be understood as follows. For $\tau_X \stackrel{<}{\sim}10^{-10}$ sec, the $X$ particles decay before reaching the detector ($e^{-d/\tau_X \gamma_X}\ll 1$). For $\tau_X \stackrel{>}{\sim}10^{-10}$ sec, the $X$ particles survive up to the detector ($e^{-d/\tau_X \gamma_X}\simeq  1$) and the number of events is determined by $\Lambda^{-2} \tau_X^{-1}$ so the curve follows a $\tau_X \Lambda^2=cte$ behavior.  We have also examined the bound that can be derived by FASER during the LHC run III. Since the volume of FASER is smaller than FASER$\nu$, the bound shall be weaker by a factor of 1.5. For comparison, we also show the lower bound that can be obtained during the high luminosity run of the LHC with an integrated luminosity of 3000 fm$^{-1}$
by FASER2 setup with a length of 5~m and a radius of 1~m.   If approved, FASER2 will collect data  during 2026-2035.  The region between the two horizontal lines corresponds to the viable region for the ``Axion Quality Problem" at axion ($X'$) mass of 3 GeV according to \cite{Hook:2019qoh}.

As seen from Fig. 2, for $10^{-10} {\rm sec}< \tau_X < {\rm few }\times 10^{-5}{\rm sec}$, a range of $\Lambda$ which is still allowed can be probed by FASER$\nu$ during run III  of the LHC. The strongest bound can be obtained for $\tau_X \sim 10^{-9}$ sec which
is $\Lambda> {\rm few }\times 10^5$~GeV. FASER2 can improve this lower bound
to $\Lambda >{\rm few }\times 10^7$ GeV and can probe the yet-unconstrained range of $\Lambda$ up to $\tau_X \sim 0.1$ sec.

In sect. \ref{production}, we have found that the number of $X$ particles emitted in the direction of  the detector will be $\sim 10^5\times (2 \times 10^4~{\rm GeV}/\Lambda)^2$. The number of the $X$ particles decaying at the detector is given by Eq. (\ref{NomEvents}) which in the favorable range of $\tau_X \gamma_X \sim d=480$~m will be suppressed by a factor of $s_z/d \sim 3 \times 10^{-3}$. Thus, we may estimate the number of events at FASER$\nu$ to be $\sim {\rm few}\times 100\times (2 \times 10^4~{\rm GeV}/\Lambda)^2$.


\section{Resolving dark sector decay chain by FASER$\nu$ \label{chain}}
Let us now discuss to what extent FASER$\nu$ can help to unveil a rich structure in the dark sector by identifying the intermediate steps of $X$ decay chains. When we open up the possibility of chain decay, the possibilities become endless. As a representative case, we focus on a scenario in which $X \to Y \eta \bar \eta$ in the first step and then $\eta \to l\bar{l}$, where
$X$, $Y$, and $\eta$ are all neutral particles and $\eta$ lives long enough to travel a distance larger than $\sim 1$ mm before decay. As we shall see in sect. \ref{ConnectionTODM}, $Y$ can be a suitable dark matter candidate whose abundance is set by the freeze-out scenario with annihilation $Y +\bar{Y} \to \eta +\bar \eta $.
To illustrate the capacity of FASER$\nu$ to resolve the decay chain of $X$, we study how well FASER$\nu$ can disentangle our scenario described above from the following scenarios for the $X$ decay at the detector:
1) $X \to \eta \bar \eta$ and subsequently $\eta \to l\bar{l}$ with $\eta$ traveling a distance larger 1 mm and smaller than the detector size before decay;
2) $X \to l \bar{l} l \bar{l}$ which may correspond to $X \to \eta \bar \eta$ with $\eta$ promptly decaying into $l \bar{l}$; 
3) $X\to l \bar{l}$.

Let us first focus on the scenario $X \to Y \eta \bar \eta$ and  $\eta \to l\bar{l}$.
The decay of $\eta$ can proceed through a coupling of form
\be \lambda_\eta \eta \bar{l} l
\label{lambda-eta} . \ee The lifetime will then be $\tau_\eta =4\pi/(m_\eta \lambda_\eta^2)$. For $m_\eta<m_\mu$, $\eta$ can only decay into $e^- e^+$. To  guarantee that the $\eta$ particles which are produced inside the FASER$\nu$ detector decay before leaving the detector, $\lambda_\eta$ should be larger than 
$10^{-7}$. (That is for $\lambda_\eta > 10^{-7}$,  $\tau_\eta E_\eta /m_\eta < 1$ m.) If the decay length is so short, the electron beam dump experiments such as E137 \cite{E137} cannot constrain because $\eta$ decays before reaching the detector. Moreover, the distance travelled by $\eta$ before decay will be too short to affect the supernova evolution or its cooling process.
{ In sect. \ref{ConnectionTODM}, we discuss how such coupling can be obtained in a consistent way.}

\begin{figure}[h]
\hspace{0cm}
\includegraphics[width=0.8\textwidth, height=0.4\textwidth,trim={0cm 16cm 4cm 2cm},clip]{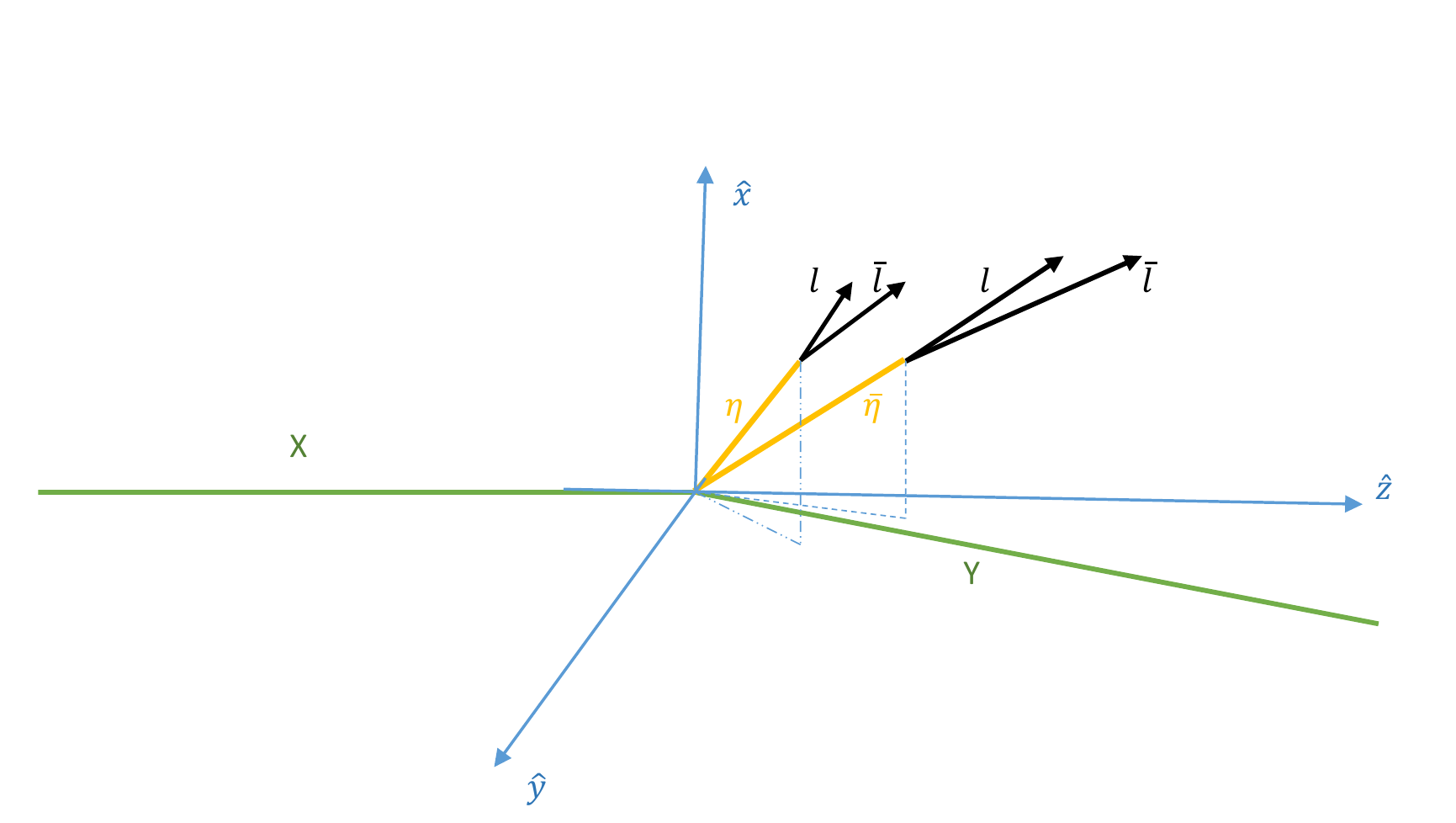}
\caption[...]{
Schematic demonstration of the $X$ decay into $Y$, $\eta$ and $\bar{\eta}$ and the subsequent production of
 charged lepton pairs from the $\eta$ and $\bar{\eta}$ decays. $\hat{z}$ is along the beam direction ({\it i.e.,} along the line connecting FASER$\nu$ to the IP).
\label{X_decay}
}
\end{figure}

\begin{figure}[h]
\hspace{0cm}
\includegraphics[width=0.65\textwidth, height=0.4\textwidth,trim={0cm 0cm 0cm 0cm},clip]{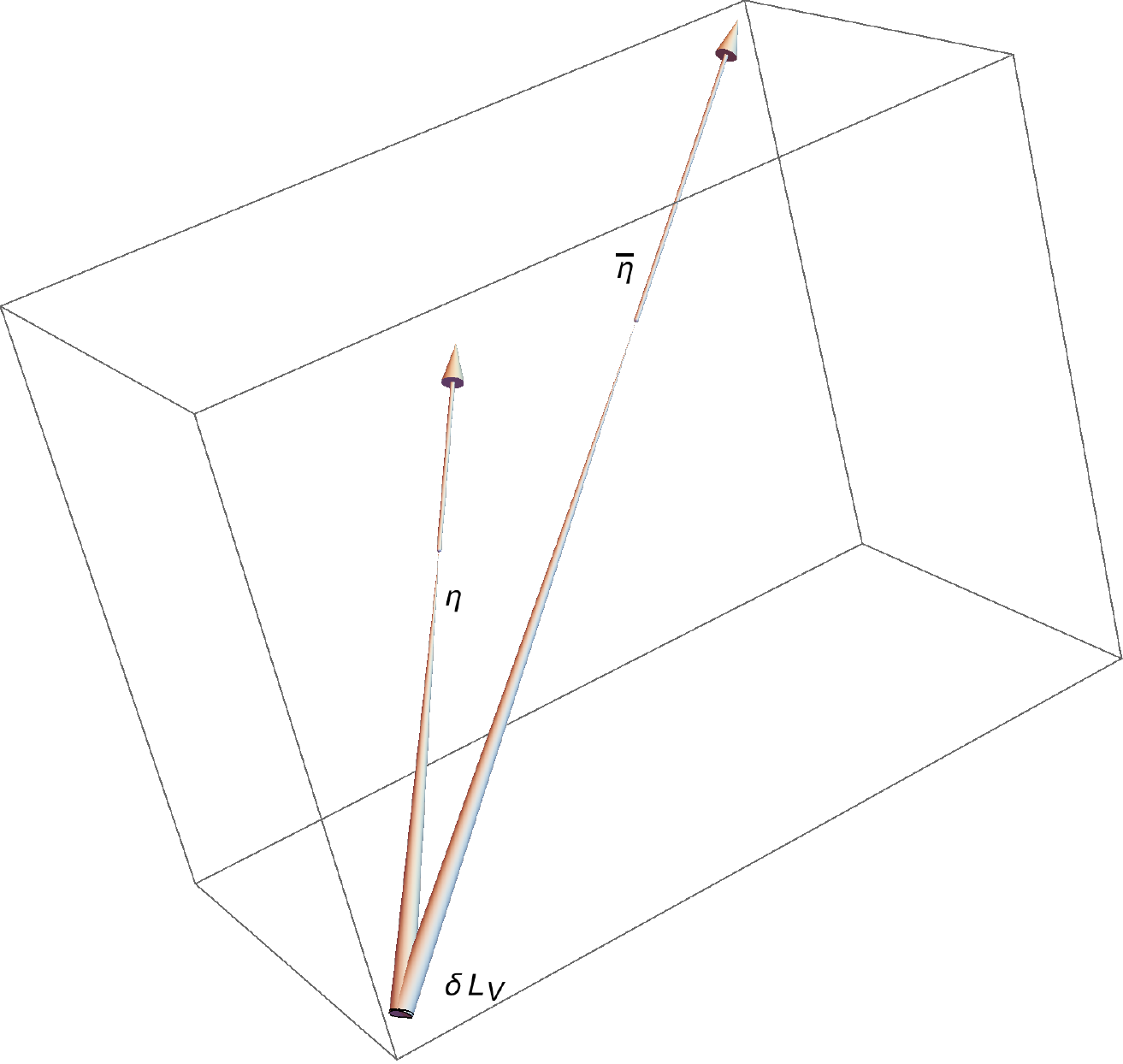}
\caption[...]{
Schematic demonstration of $\eta$ and $\bar{\eta}$ trajectories from decay of $X$. The arrows show the directions of the momenta of $\eta$ and $\bar{\eta}$ which can be reconstructed by measuring the momenta of the final lepton pairs. The narrow cones show the reconstructions of the trajectories of $\eta$ and $\bar{\eta}$, taking into account the uncertainties in determination of the momenta of $\eta$ and $\bar{\eta}$.  The opening angles  of the narrow cones are characterized by $\Delta \theta_\eta$. The $X$ decay vertex lies within the region where the two narrow cones intersect. The region  is denoted by $\delta L_V$. 
\label{etas}
}
\end{figure}

Fig.~\ref{X_decay} schematically shows the trajectory of an $X$ particle which enters from one side of the detector and those of its decay products. The trajectories of $X$ and $Y$ (which both are odd under a $Z_2$ symmetry stabilizing the dark matter candidate, $Y$) are shown by green lines. $Y$ is stable and leaves the detector. The $\eta$ and $\bar{\eta}$, whose trajectories are shown by yellow lines, decay into the $l \bar{l}$ pairs within the detector. The average angle between the $\eta$ particles with each other and the forward direction is of order of $\theta_\eta \sim ({m_X}/{E_X})\left( 1-4m_\eta^2/(m_X-m_Y)^2\right)^{1/2}$. The $X$, $Y$, $\eta$ and $\bar{\eta}$ particles are all neutral and cannot be seen. Only the trajectories of the final charged leptons can be observed. As we discuss below, FASER$\nu$ can resolve the vertices at which $\eta$ and $\bar{\eta}$ decay  with a remarkable precision. The momenta of $\eta$ and $\bar{\eta}$ can be also derived. If we know the direction of the momenta of $\eta$ and $\bar{\eta}$ with a precision, $\Delta \theta_\eta$, the trajectories of $\eta$ and $\bar{\eta}$ can be reconstructed within an uncertainty cone whose opening angle is characterized with $\Delta \theta_\eta$.
The $X$ vertex lies within the small region where these two cones intersect.  We denote the intersection of the two narrow cones with $\delta L_V$.
The smaller the $\delta L_V$, the smaller the uncertainty in determination of the $X$ decay vertex.  This has been schematically demonstrated in Fig.~\ref{etas}.

The typical angular separation of the two leptons emitted from $\eta$ decay (as well as the angle that each lepton makes with the direction of the momentum of the parent $\eta$) is given by $\theta_l\sim [m_\eta^2-4m_l^2]^{1/2}/E_\eta$. At FASER$\nu$, the vertex of each $\eta$ decay can be constructed with a remarkable precision of $\sigma_{pos}=0.4~ \mu$m \cite{Abreu:2019yak}. Considering that the decay length
is a stochastic parameter, $\eta$ and $\bar{\eta}$ coming from the decay of a single $X$ will travel different distances before decay. {In fact,  the decay vertices of $\eta$ and $\bar{\eta}$ will be separated by a distance of order of the decay length itself. Let us suppose this separation is larger than the size of the electron positron shower in tungsten (which, including the emulsion layers, will be smaller than 1.5~cm \cite{Abreu:2019yak}).}  It will then be feasible to determine which pair of leptons originate from a single vertex, even if the two $\eta$ particles are emitted very close to the forward direction {({\it i.e.,} even if  $\theta_l \sim \theta_\eta$ or $\theta_l >\theta_\eta$). Moreover, the   decay vertices of $\eta$ and $\bar{\eta}$ can be discerned.} This feature has been also demonstrated in Fig.~\ref{X_decay}.  
{In case $l_{\eta},l_{\bar{\eta}}<1.5$~cm, the separation of the vertices will be more challenging as one vertex may be located inside the electromagnetic shower originated  by the particles from the other vertex.}

In any case if $E_\eta \gg m_\eta$, the angle between the pair of leptons from $\eta$ and $\bar\eta$ decays, $\theta_l$,  will be very small. This has an immediate (and ironically positive) consequence for the reconstruction of the direction of the momentum of $\eta$. FASER$\nu$ has an excellent angular resolution of reconstructing the direction of tracks. However, the energy resolution is poor and only at the level of $\Delta E/E \sim 30 \%$ \cite{Abreu:2019yak}. If the angle between $l \bar{l}$ pair is small, it will still be feasible to reconstruct the direction of $\eta$ with an accuracy of $$\Delta \theta_\eta \sim \sqrt{(\Delta E/E)^2\theta_l^2/2+ 2\sigma_{ang}^2},$$ where $\sigma_{ang}$ is the angular resolution of the track reconstruction $\sigma_{ang}=\sqrt{2}
\sigma_{pos}/L_{tr}$ where $\sigma_{pos}=0.4~\mu m$ and $L_{tr}$ is the track length of the charged leptons. Taking $m_\eta \sim 100$ MeV and $E_\eta \sim$ few 100 GeV, $\Delta \theta_\eta \sim {\rm few}\times 10^{-4}$. As mentioned before, the position of the vertex in which the charged tracks meet can be reconstructed with a precision of $\sim 0.4~\mu$m. Then, the ``invisible track" of $\eta$ and $\bar{\eta}$ can be reconstructed. If these tracks meet within the uncertainties, it will be an indication that they are coming from the decay of the same particle and are not a pile-up from two separate events.
As long as $m_\eta,m_Y \ll m_X$, the angular separation of $\eta$ and $\bar \eta$ will be of order of $\theta_\eta= m_X/E_X$ for both $X\to \eta \bar{\eta}$ and $X\to Y\eta \bar{\eta}$.
The typical angle between the direction of $\eta$ (and that of $\bar{\eta}$) and the direction of the parent $X$ is also of order of $\theta_\eta$.
Let us denote the distance traveled by $\eta$ and $\bar{\eta}$ before decay by $l_\eta$ and $l_{\bar{\eta}}$ and take them to be larger than 1 mm. { The point at which $\eta$ and $\bar{\eta}$ tracks meet can be reconstructed with an uncertainty of
\be \delta L_z\sim [(l_\eta^2+l_{\bar{\eta}}^2)(\Delta \theta_\eta/\theta_\eta)^2+2 \sigma_{pos}^2/\theta_\eta^2]^{1/2}\ee in the forward direction and  uncertainties of \be \delta L_x\sim \delta L_y \sim [(l_\eta^2+l_{\bar{\eta}}^2)(\Delta \theta_\eta)^2+2 \sigma_{pos}^2]^{1/2}\ee in the vertical directions. Notice that $\delta L_x,\delta L_y\ll \delta L_z$ so the uncertainty is elongated in the forward direction as expected. Since $\Delta \theta_\eta \ll \theta_\eta \sim m_X/E_X$ and $\sigma_{pos}/\theta_\eta \ll 1~{\rm mm} < l_\eta, l_{\bar{\eta}}$, we find $\delta L_x,\delta L_y\ll \delta L_z \ll l_\eta,l_{\bar{\eta}}$ so the $X$ decay vertex can be resolved with reasonable precision. 
Taking $E_\eta\sim E_X/3$ and $m_e \ll m_\eta \ll m_X$, we find $\delta L_z\sim m_\eta/m_X l_\eta$.
}

 As long as $\delta L_z \ll l_\eta,l_{\bar{\eta}}$, even a handful of events will be enough to discriminate between this scenario and the ones in which $X\to l \bar{l}l \bar{l}$ or $X\to l \bar{l}$. Let us suppose FASER$\nu$ registers two pairs of $l \bar{l}$ from the decays of two separate $X$ via $X\to l\bar{l}$. The track of
$X$ can be reconstructed with a precision of $[ (\Delta E/E)^2(m_X/E_X)^2/2+\sigma_{ang}^2]^{1/2}$. Considering that the typical separation of the $X$ vertices will be $\sim {\rm few ~ cm}-{\rm few ~10~ cm}$, the reconstructed tracks will meet only far outside the detector around IP where the two $X$ particles are produced from $X'$ decay. Thus, it can be discriminated from our scenario that includes the intermediate $\eta$ and $\bar{\eta}$ coming from a $X$ decay vertex inside the detector.

Let us now discuss whether FASER$\nu$ can distinguish between $X\to \eta \bar{\eta}$ and $X\to Y\eta \bar{\eta}$.
Considering the three-dimensional nature of the three-body decay $X\to \eta \bar{\eta } Y$, by reconstructing the momenta of $\eta$ and $\bar{\eta}$, it would be possible to determine if the decay of $X$ is a two-body or three-body decay. In other words, if we can reconstruct the direction of the projection of the $\eta$ and $\bar{\eta}$ momenta onto the plane perpendicular to the direction of the incoming $X$ particles (the direction of the line connecting IP to the $X$ decay vertex), we can check whether a third neutral particle should have been emitted in the $X$ decay to balance the transverse momentum.
An alternative, but not an independent way, to test for the emission of the $Y$ particles along with $\eta$ and $\bar{\eta}$ is to reconstruct the invariant mass of the two pairs of $l^-l^+$. If $X\to \eta \bar{\eta}$, this invariant mass should be equal to $m_X$. Thus, the distribution of the invariant mass of the two $l^-l^+$ pairs should be monochromatic but for $X\to \eta \bar{\eta}Y$, this distribution will be continuous.
Regardless of the method invoked, we have to be able to measure the transverse momenta of the $\eta$ particles. This of course requires relatively large transverse momentum. For a given $m_X$, the highest traverse momenta can be achieved for $m_\eta, m_Y \ll m_X$. If we take $m_X=1$ GeV, this condition implies $m_\eta <2 m_\mu$ so the only possible decay mode for $\eta$ will be the decay into the electron-positron pair. For heavier $m_X$ and $ m_\eta$, there is a possibility of $\eta$ decay into muon pair.

The typical traverse momenta of the $\eta \bar{\eta}$-pair ($|\vec{p}_\eta^t +\vec{p}_{\bar\eta}^t|$) in the $X \to \eta \bar{\eta} Y$ for $m_\eta,m_Y \ll m_X$ will be of order of $\sim m_X/2$ and  can be reconstructed with a precision of order of
\be \label{DeltaPt}\Delta P_t=\left(2 (\frac{\Delta E}{E})^2 [(p_\eta^t)^2+(p_{\bar\eta}^t)^2]+ (\sum |\vec{p}_l|^2+\sum |\vec{p}_{\bar{l}}|^2) \sigma_{ang}^2+|\vec{p}_\eta +\vec{p}_{\bar\eta}|^2 (\Delta \hat{z})^2
\right)^{1/2}\ee where $(\sum |\vec{p}_l|^2+\sum |\vec{p}_{\bar{l}}|^2)\sim |\vec{p}_\eta|^2\sim |\vec{p}_{\bar\eta}|^2$
is the sum of the squares of the momenta of all final leptons. In Eq. (\ref{DeltaPt}),
the first term proportional to $(\Delta E/E)^2 \sim 0.3^2$ is due to the uncertainty in the lepton momentum measurement, the second term is due to the uncertainty in the measurement of the direction of the four final leptons and finally $\Delta \hat{z} \sim 10^{-3}$ is the uncertainty in the alignment of the detector relative to the beamline.
It is straightforward to confirm that the first term will dominate. To determine whether there is missing traverse momentum, $\Delta P_t$ must be smaller than the sum of the traverse momenta of $\eta$ and $\bar{\eta}$: $\Delta P_t<|\vec{p}^t_\eta+\vec{p}^t_{\bar\eta}|$. For the events that the projections of the momenta of $\eta$ and $\bar{\eta}$ onto the $x-y$ plane make an angle smaller than 90$^\circ$, this condition is readily satisfied. As we saw, for $\Lambda$ close to the lower bound $\Lambda \sim 10$ TeV, the number of events will be $O(100)$ so there will be enough statistics to determine whether a third invisible state along with $\eta$ and $\bar{\eta}$ is emitted in the $X$ decay or not. 


The invariant mass of the pair of leptons that meet at a vertex gives the $\eta$ mass: $m_\eta^2= (p_l +p_{\bar{l}})^2$.
By studying the distribution of $m_\eta^2$ measured for the observed pairs, we can test our assumption that $\eta$ decays to a lepton pair rather than going through a three-body decay.
{As we shall discuss below, reconstructing $m_\eta$ is also a powerful
tool to veto the background from the $e^-e^+$ pair production from the gamma produced by the neutrino beam interaction at the detector.}
With a single pair, the precision in the measurement of $m_\eta$ is $\Delta m_{sin}^2/m_\eta^2\sim [2(\Delta E/E)^2+2 (\sigma_{ang}/\theta_l)^2]^{1/2} $ which cannot be better than $40 \%$. However, if $N$ pairs are detected, the precision will be improved by $\Delta m_{sin}^2/\sqrt{N}$.  Remember that $N\sim {\rm few}\times 100 \times (2 \times 10^4~{\rm GeV}/\Lambda)^2$. As a result, the precision in the $\eta$ mass will be at
the level of a few percent for $\Lambda \sim 10^4$ GeV to a few ten percent for $\Lambda \sim 10^5$ GeV.
Notice that in our scenario $N$ should be an even number as for any lepton antilepton pair coming from an $\eta$, there is another pair coming from $\bar{\eta}$.

As discussed before, $1.5~{\rm cm}<l_\eta, l_{\bar{\eta}}< {\rm few }\times 10$~cm is the range  within which the values of $l_\eta$ and $l_{\bar{\eta}}$ can be extracted from the FASER$\nu$ data. For smaller values of $l_\eta, l_{\bar{\eta}}$, the uncertainty in the determination of $l_\eta$ ({\it i.e.,} $\sqrt{2} \sigma_{pos}/\theta_\eta$) will be larger than $l_\eta,l_{\bar{\eta}}$ or one vertex may stand within the electromagnetic shower caused by the other vertex. On the other hand, for $l_\eta, l_{\bar{\eta}}>$1~m, the $\eta$ decay takes place outside the detector, leaving no signal. For $l_\eta \sim s_z/3-s_z/2$, the chances that the whole chain of $X\to \eta \bar{\eta}Y$, $\eta \to e^-e^+$ and $\bar{\eta}\to e^-e^+$ takes place inside the detector is slim. The $X$ particles decaying at a distance of $\sim l_\eta$ before reaching the detector can lead to two pairs of $e^-e^+$. The reconstructed tracks of the associated $\eta$ and $\bar{\eta}$ shall meet outside the detector but it is still possible to extract $l_\eta$ and $l_{\bar{\eta}}$ for such events. $\eta$ and/or $\bar{\eta}$ from particles decaying in the second half of the detector may undergo decay after leaving the detector. In this case, we cannot derive $l_\eta$ and $l_{\bar{\eta}}$. If the statistics is high ({\it i.e.,} if $\Lambda$ is small), it will still be possible to derive information despite $l_\eta\sim s_z/3-s_z/2$.

 For $1.5~{\rm cm}<l_\eta,l_{\bar{\eta}}<$ few 10 cm, we can extract the lifetimes of $\eta$ and $\bar{\eta}$ from the measured $l_\eta$ and $l_{\bar{\eta}}$. The mean value of $\tau_\eta$ can be written as
\be \bar{\tau}_\eta = \frac{\sum_{i=1}^N l_\eta^i m_\eta/E_\eta}{N}\ee
which maximizes the probability of decaying of the $N$ pairs of $\eta$ and $\bar{\eta}$ after traversing the measured $l_\eta^i$:
\be P(\tau)\propto \prod_i \left(\exp^{-\frac{l_\eta^i m_\eta}{\tau E_\eta}} \frac{1}{\tau}\right) \ee
where $E_\eta =E_l +E_{\bar{l}}$. As discussed above, the systematic error in the measurement of $l_\eta$ and $l_{\bar{\eta}}$ can be neglected. A major error comes  from the uncertainties in the extracted $m_\eta$ and $E_\eta$ which should be of order of $ 40 \% /\sqrt{N}$. Let us now discuss the statistical uncertainty in extracting the lifetime of $\eta$. Let us define $(\delta \tau)^{stat} $ such that $P(\bar{\tau}\pm (\Delta \tau)^{stat})/P(\bar{\tau})=1/2$. With this definition
$$\frac{\delta \tau^{stat}}{\bar{\tau}}=(\frac{2 \log 2}{N})^{1/2}$$ which will dominate over the  error from the uncertainties of $m_\eta$ and $E_\eta$.
As we discussed before, for $\Lambda \sim$10 TeV and $\tau_X \gamma_X \sim d$, we expect $N\sim 100$ so the precision in the determination of the lifetime can be better than 10 \%.

	Let us now discuss the limitations caused by uncertainties and the background in determinibg the parameters of the model. Since we a priori do not know the momenta of the initial $X$, we cannot determine the masses of $X$ and $Y$ on event by event basis. However, at the statistical level, some information should be extractable by studying the energy distribution of the events. The information will however suffer from the uncertainty on the Parton Distribution Function (PDF). The same uncertainty also limits the accuracy of determining the $X'$ coupling to gluons and the $X$ lifetime as shown in Fig.~\ref{lower_bound}.
	However, determining the $\eta$ mass and lifetime is free from the PDF uncertainties.
	For $\eta \to e^-e^+$, the signature of our scenario is two pairs of $e^-e^+$ moving in a direction close to that of the beam plus missing transverse momentum. At FASER$\nu$, this signature can be mimicked by the $\pi^0$ produced in the scattering of the neutrino beam coming from the IP off the detector. That is  because $\pi^0$ decays into a gamma pair which in turn leads to two $e^- e^+$ pair productions. This background can be vetoed by reconstructing the invariant mass of both $e^-e^+$ pairs as well as the invariant mass of the two pairs combined. While the invariant masses of the $e^-e^+$ pairs in our scenario should be equal to $m_\eta$ ({\it i.e., }$({p}_{e^-}+{p}_{e^+})^2=m_\eta^2$), the invariant masses of the pairs from the background should vanish ({\it i.e., }$({p}_{e^-}+{p}_{e^+})^2\simeq m_\gamma^2=0$). Moreover, while the invariant mass of the two pairs combined in our scenario is distributed around $\sim (m_X -m_Y)\sim$GeV, that from the background is monochromatic and equal to $m_{\pi^0}=135$~MeV. By reconstructing these invariant masses, the background can be significantly reduced.  	Moreover, the majority of neutrino beam induced events will be charged current interactions  which can be identified from our signal by the reconstruction of the associated  charged leptons. As a result, only $\pi^0$ from neutral current neutrino events can count as a background to our signal. The number of such events will be around 4000 in the run III. Moreover, in these events, multiple pions can be created which will be another handle to further reduce the background. Unlike FASER$\nu$, the FASER experiment is hollow so there will be no neutrino beam induced background altogether.
	We have therefore neglected the background in our analysis.

\section{Connection to dark matter\label{ConnectionTODM}}
In previous sections, we have  introduced singlet particles $X$, $Y$ and $\eta$ with masses of  few 100~MeV to few GeV such  that $X\to Y \eta \bar{\eta}$. Here, we show that $Y$ can provide a suitable dark matter candidate within the freeze-out scenario for dark matter production in the early universe with $\langle \sigma (Y\bar Y \to \eta \bar{\eta}) v\rangle\sim 1$ pb \cite{Steigman:2012nb}.
  $\eta$ plays the role of the intermediate particle which eventually decays into the standard model particles. Since $\eta$ is heavier than 20 MeV and decays with a lifetime much shorter than $10^{-8}$ sec, it can avoid the bounds from BBN. A $Z_2$ symmetry, under which only $X$ and $Y$ are odd, guarantees the stability of the lighter one which is taken to be $Y$.

The production of the dark sector in the early universe can proceed via two channels. For $T>m_{X'}$ and $\Lambda\stackrel{<}{\sim} 10^8$ GeV, $X'$ can be produced with a rate of $O(T^3/\Lambda^2)\gg H=T^2/M^*_{Pl}$ at $T\stackrel{>}{\sim} m_{X'}$. The decay of $X'$ can create $X$ and then the subsequent decay of $X$ creates $Y$ and $\eta$. The $\lambda_\eta$ coupling between $\eta$ and the $e^-e^+$ pair with $\lambda_\eta \stackrel{>}{\sim} 10^{-6}$ can keep $\eta$ in equilibrium with the plasma down to $T\sim m_\eta$. The couplings of $\eta$ to $X$ and $Y$ can keep $\eta$ in equilibrium with these particles as long as the temperature is above their masses.
We require $\langle\sigma(Y+\bar{Y} \to \eta +\bar{\eta}) v \rangle\sim 1$ pb in order to have a successful freeze-out 
scenario \cite{Steigman:2012nb}.
In the following,
we discuss two cases where $X$ and $Y$ are Dirac fermions or scalars one by one.
\begin{itemize}
\item {\textit{Fermionic $X$ and $Y$}:} We introduce a new heavy fermion $V$ which is singlet under the SM and odd under the $Z_2$ symmetry stabilizing $Y$. In addition to this exact $Z_2$ symmetry, let us also impose an approximate $Z_2$ symmetry under which only $\eta$ and $V$ are odd. This symmetry forbids terms such as 
$  \eta \bar{Y}Y$,  $\eta \bar{X}X$, $ \eta \bar{V}V$ and more importantly $\eta \bar{X}Y$ or $\eta \bar{Y}X$  which in turn forbids $X\to \eta Y$ and $X\to \bar{\eta} Y$.  This symmetry allows \be g_{Y} \eta \bar{Y}V+g_X \eta \bar{X}V+ {\rm H.c.}\ee
In case that $\eta$ is a real scalar, the $t$ and $u$ channel contributions
cancel each other. If $\eta$ is a complex scalar lighter than $Y$,
the annihilation cross section of $Y$ can be written as
$$\langle \sigma (Y +\bar{Y} \to \eta +\bar\eta) v \rangle= \frac{g_Y^4}{32\pi m_V^2}.$$
Taking $\langle \sigma (Y +\bar{Y} \to \eta +\bar\eta)v \rangle\sim $ pb, we find $g_Y \sim 0.07 \times (m_V/10~{\rm GeV})^{1/2}$.

For $m_Y\ll m_X\ll m_V$, the decay rate of the $X$ particle is given by the following relation
$$\Gamma(X\to Y \eta \bar{\eta})\sim \frac{g_{X}^2g_{Y}^2}{100 \pi^3 m_V^2}m_X^3.$$
To obtain $\Gamma(X\to Y \eta \bar{\eta})(m_X/E_X) \sim (480~{\rm m})^{-1}$, we should have
$$g_X \sim 10^{-4}~ (m_V/10~{\rm GeV})^{1/2} (E_X/ 500~{\rm GeV})^{1/2}({\rm GeV}/m_X)^2.$$
The long lifetime of $X$ requires $g_X\ll g_Y$. Notice that in the limit of a symmetry under which $X$ is even but $\eta V$ and Y are odd, $g_X$ should be zero. Thus, such approximate symmetry justifies the hierarchy between $g_X$ and $g_Y$. 

\item {\textit{Scalar $X$ and $Y$}:} Let us  consider the following quartic couplings between these scalar fields:
\be \lambda_1 \bar Y X \bar \eta \eta+\lambda_2 \bar X X \bar \eta \eta
+\lambda_3 \bar Y Y \bar \eta \eta
. \label{quartic} \ee
$\lambda_1$ is the coupling that leads to the $X$ decay and must be nonzero:
\be \Gamma (X\to Y \bar{\eta \eta})\sim \frac{\lambda_1^2m_X}{100 \pi^3}.\ee
To obtain $\Gamma (X\to Y \bar{\eta \eta}) (m_X/E_X)\sim 1/(480 ~m)$, we should have $\lambda_1 = 8 \times 10^{-7} ({\rm GeV}/m_X) (E_X/500~ {\rm GeV})^{1/2}$. If we take
$ \langle \sigma ( Y \bar Y\to \eta \bar \eta)v \rangle \sim \lambda_3^2/(4\pi m_Y^2)\sim 1 ~ pb$, which implies $\lambda_3 \sim 10^{-4} \sqrt{m_Y/{\rm GeV}}$, we shall have the standard freeze-out scenario. 
Unlike the case for fermionic $X$ and $Y$, $\eta$ can be either real or complex.

We have taken the lifetime of $X$ to be $\tau_X= (E_X/m_X)(480~ {\rm m}/c)=8 \times 10^{-4}~{\rm sec} (E_X/500~{\rm GeV})({\rm GeV}/m_X)$. This means the $X$ particles will decay at a temperature of $\sim$100 MeV which is well below the $X$ mass.
To avoid the consequences of out of equilibrium decay of $X$,  we may take $\lambda_2$ large enough to lead to sufficient annihilation of $X \bar X$ pair to $\eta \bar \eta$: $\lambda_2 \gg 10^{-4} \sqrt{m_X/{\rm GeV}}$. Thus, we find that within the freeze-out scenario $\lambda_2 \gg \lambda_3 \gg \gg \lambda_1$. The hierarchy between $\lambda_1$ and $\lambda_3$ can be explained by an approximate $Z_2\times Z_2$ symmetry which is broken by $\lambda_1$ to a $Z_2$. The residue $Z_2$ symmetry is the same symmetry which keeps $Y$ stable. Eq. (\ref{quartic}) enjoys yet another $Z_2$ symmetry under which only $\eta$ is odd. Without this extra $Z_2$ symmetry, we could have trilinear terms such as $\bar{X} Y \eta$ that leads to two-body decay of $X\to Y \eta$ which we intend to avoid.
\end{itemize}
 Notice that in both cases enumerated above, $\lambda_\eta$ breaks the approximate $Z_2$ symmetry under which  $\eta$ is odd. This approximate symmetry explains the smallness of $\lambda_\eta$ in Eq. (\ref{lambda-eta}) and therefore the longevity of $\eta$.
Longevity of $X$ is explained by the approximate $Z_2 \times Z_2$ symmetry that is broken to the exact $Z_2$ symmetry that protects $Y$  (DM) against decay.

{Let us now discuss how the $\lambda_\eta$ coupling in Eq. (\ref{lambda-eta}) can be induced after electroweak symmetry breaking.  If $\eta$ is real, this coupling can be obtained by a mixing between $\eta$ and the SM Higgs. Such a mixing can  in priniciple originate from a trilinear coupling of form $\eta H^\dagger H$ but a few challenges should be addressed: 1) After electroweak symmetry breaking, apart from a mixing between $H$ and $\eta$, this term also induces a linear term for $\eta$ that leads to a large VEV for $\eta$. If $\eta$ gets a VEV, the new fermions also mix and we have to revisit all the above discussion. 2)
	As we showed in sect  \ref{dark-sector}, in order for   $\eta$ to decay inside the detector and give rise to an observable signal, $\lambda_\eta$ should be around or larger than $10^{-7}$.
	A coupling  of $10^{-7}$ to the electrons can be obtained through a mixing of order of $\lambda_\eta (\langle H\rangle /m_e) \sim 0.1$ to the SM Higgs. To obtain a shorter lifetime,  $\lambda_\eta$ must be larger. Both these challenges can be avoided by introducing an inert Higgs doublet, $\Phi$, with arbitrary coupling to $l \bar{l}$ pair and by mixing its neutral coupling with $\eta$. The mixing can be obtained from the trilinear coupling of $\eta (H^\dagger \Phi+\Phi^\dagger H)$. Such scenarios have been extensively studied in the literature so we shall not elaborate further. For complex $\eta$, the SM Higgs can only mix with a linear combination of Im$[\eta]$ and Re$[\eta]$. The other linear combination perpendicular to it will not obtain effective coupling to the $l \bar{l}$ pair. Like the case of real $\eta$, by introducing an inert Higgs doublet, whose neutral component is complex, this problem can be solved. The mixing can come from the trilinear term $\eta H^\dagger \Phi+ \eta^\dagger \Phi^\dagger H$.}


\section{Concluding remarks\label{summary}}

We have studied the capabilities of the upcoming FASER$\nu$ experiment in the quest for light dark sector. In order to produce the dark particles at the LHC, there has to be an intermediate state, $X'$, that couples to the partons. We have discussed three different possibilities for such intermediate state: (1) a light gauge boson coupled to quarks; (2) a light scalar with a Yukawa coupling to the quarks; (3) an axion-like particle coupled to a gluon pair with a coupling ($1/\Lambda$) independent of its mass as in Ref. \cite{Hook:2019qoh}. We have focused on the third option which is also theoretically motivated by the ``axion quality problem." Generally, if the coupling of the intermediate particle is large, the production of $X'$ will be significantly abundant but it will immediately decay back to the partons before having time to reach the FASER$\nu$ detector which is located 480 m away. In our model, the main decay mode of $X'$ is into a pair of new neutral particles $X$ and $\bar{X}$.  The lifetimes of $X$ and $\bar{X}$ are relatively large such that they can survive up to the position of FASER$\nu$.
We have computed the spectrum of the $X$ particles that are emitted towards FASER$\nu$. We have shown that FASER$\nu$ during  LHC run III and then FASER$2$ during the high luminosity run of the LHC can probe a wide range of  $\Lambda$ that is still unexplored.


It is well-known that if the dark matter mass is lower than $\sim 10$ GeV, it should have interactions other than the electroweak interactions in order for the freeze-out dark matter production scenario to lead to the observed abundance in the universe. In a class of models, the light dark matter pair, $Y$ and $\bar{Y}$, annihilate to a pair of intermediate new neutral particles, $\eta$ and $\bar{\eta}$, which eventually decay into the standard model particles. We have found that 
FASER$\nu$ with its superb track reconstruction capabilities is an ideal setup to test such scenarios. We have elaborated on the model-building aspects of this wide class of models where the $X$ and $Y$ particles can be either scalar or fermion. In our models, $X$ decays to dark matter $Y$ and a pair of $\eta \bar{\eta}$ which in turn decay into two pairs of $e^-e^+$.
The signature of the event will be therefore two pairs of $e^- e^+$ meeting in two different vertices. Such a signature is practically free from the background but there is yet another signature for the events as described below. By measuring the momenta of the final charged leptons, the four momenta of  $\eta$ and $\bar{\eta}$ can be reconstructed. Moreover, FASER$\nu$ can resolve the vertices of $\eta$ and $\bar{\eta}$ decay with a remarkable precision. As a result, the tracks of $\eta$ and $\bar{\eta}$ can be reconstructed. The tracks have to meet in the $X$ decay vertex, providing yet another signature. We have discussed the precision of determining the $X$ decay vertex. If the statistics allow,  their lifetime can be extracted by measuring the decay length of $\eta$ and $\bar{\eta}$. In the favorable range of $\Lambda$  and $\tau_X$, the precision of determining the lifetime can be at the level of a few percent.

Measuring the four momenta of the final charged leptons, the invariant mass of $e^- e^+$ pairs can be reconstructed and the mass of $\eta$ can be extracted. Again the precision can be as good as percent level. Moreover, it will be possible to distinguish this model from a model in which $\eta$ goes through a three-body decay by studying the distribution of the invariant mass of the final lepton pairs. By reconstructing the transverse momenta of $\eta$ and $\bar{\eta}$ for each $X\to Y \eta \bar{\eta}$, the emission of the invisible particle $Y$ can be tested. The signature of the dark matter $Y$ particles is missing transverse momentum. We have quantified the conditions that have to be met to discriminate between the two scenarios in which $X\to \eta \bar{\eta}$ and $X\to Y\eta \bar{\eta}$.

We have  discussed that the majority of $X^\prime$ particles directed towards FASER$\nu$ are produced in the $2 \to 1$ processes: $gg \to X^\prime$. The  $X^\prime$ particles can also be produced via $2\to 2$ processes such as $q+g \to q+X^\prime$ or $g+g \to X^\prime+g$ \cite{Hook:2019qoh}. The $X^\prime$ particles produced via $2\to 2$ processes will typically  have large traverse momenta and can point towards the ATLAS and CMS detectors. In our model, $X^\prime$ decays to $X\bar{X}$ before  reaching these detectors. The subsequent decay of $X$ and $\bar{X}$ at CMS and ATLAS can produce two pairs of collimated $e^-e^+$ pairs. The magnetic fields at these detectors will separate the two positrons from the two electrons. Moreover, the tracks of the two electrons as well as those of the two positrons will also become separated by the magnetic field because they have different energies and therefore different Larmor radii. 
Future beam dump experiments such as SHiP and MATHUSLA will also be sensitive to this scenario.
These experiments are planned to start data taking after FASER$\nu$. Using the feedback from FASER$\nu$, these experiments can 
carry out customized searches for the new particles to extract their parameters more accurately. 

\begin{acknowledgments}
This project has received partial funding from the European Union's Horizon 2020 research and innovation programme under the Marie Sklodowska-Curie grant agreement No. 690575 (RISE InvisiblesPlus) and No. 674896 (ITN Elusives) and the European Research Council under ERC Grant NuMass (FP7-IDEAS-ERC ERC-CG 617143). YF and PB have received partial financial support from Saramadan under contract No.~ISEF/M/98223 and No.~ISEF/M/99169. YF would like also to thank the  ICTP staff and the INFN node of the InvisiblesPlus network in Padova. SP would like to thank ICTP for kind hospitality during the initial phases of this work. The authors would like to thank A. Ariga and J Feng for the useful information and the encouragement. P.B. would like to thank M. Rajaee for  useful  discussions.

\end{acknowledgments}



\begin{thebibliography}{99}

\bibitem{Ariga:2018pin}
 J.~L.~Feng, I.~Galon, F.~Kling and S.~Trojanowski,
 Phys.\ Rev.\ D {\bf 97} (2018) no.3,  035001
 doi:10.1103/PhysRevD.97.035001
 [arXiv:1708.09389 [hep-ph]];
A.~Ariga {\it et al.} [FASER Collaboration],
arXiv:1812.09139 [physics.ins-det].

\bibitem{Abreu:2019yak}
H.~Abreu {\it et al.} [FASER Collaboration],
arXiv:1908.02310 [hep-ex].
 H.~Abreu {\it et al.} [FASER Collaboration],
 arXiv:2001.03073 [physics.ins-det].
   
   \bibitem{Ariga:2018uku}
   A.~Ariga {\it et al.} [FASER Collaboration],
   Phys.\ Rev.\ D {\bf 99} (2019) no.9,  095011
   doi:10.1103/PhysRevD.99.095011
   [arXiv:1811.12522 [hep-ph]].
   \bibitem{Alimena:2019zri}
   J.~Alimena {\it et al.},
   arXiv:1903.04497 [hep-ex].
    
  \bibitem{Feng:2017vli}
  J.~L.~Feng, I.~Galon, F.~Kling and S.~Trojanowski,
  Phys.\ Rev.\ D {\bf 97} (2018) no.5,  055034
  doi:10.1103/PhysRevD.97.055034
  [arXiv:1710.09387 [hep-ph]].

\bibitem{Okada:2019opp}
N.~Okada and D.~Raut,
arXiv:1910.09663 [hep-ph].

 \bibitem{Boiarska:2019vid}
 I.~Boiarska, K.~Bondarenko, A.~Boyarsky, M.~Ovchynnikov, O.~Ruchayskiy and A.~Sokolenko,
 arXiv:1908.04635 [hep-ph].
\bibitem{Kling:2018wct}
F.~Kling and S.~Trojanowski,
Phys.\ Rev.\ D {\bf 97} (2018) no.9,  095016
doi:10.1103/PhysRevD.97.095016
[arXiv:1801.08947 [hep-ph]].
\bibitem{Helo:2018qej}
J.~C.~Helo, M.~Hirsch and Z.~S.~Wang,
JHEP {\bf 1807} (2018) 056
doi:10.1007/JHEP07(2018)056
[arXiv:1803.02212 [hep-ph]].
\bibitem{Deppisch:2019kvs}
  F.~Deppisch, S.~Kulkarni and W.~Liu,
  Phys.\ Rev.\ D {\bf 100} (2019) no.3,  035005
  doi:10.1103/PhysRevD.100.035005
  [arXiv:1905.11889 [hep-ph]].



 \bibitem{Feng:2018noy}
 J.~L.~Feng, I.~Galon, F.~Kling and S.~Trojanowski,
 Phys.\ Rev.\ D {\bf 98} (2018) no.5,  055021
 doi:10.1103/PhysRevD.98.055021
 [arXiv:1806.02348 [hep-ph]].
 \bibitem{Mohapatra:2019ysk}
 R.~N.~Mohapatra and N.~Okada,
 arXiv:1908.11325 [hep-ph].
 \bibitem{Berlin:2018jbm}
 A.~Berlin and F.~Kling,
 Phys.\ Rev.\ D {\bf 99} (2019) no.1,  015021
 doi:10.1103/PhysRevD.99.015021
 [arXiv:1810.01879 [hep-ph]].
 \bibitem{Jodlowski:2019ycu}
   K.~Jodłowski, F.~Kling, L.~Roszkowski and S.~Trojanowski,
   arXiv:1911.11346 [hep-ph].
 

 
 \bibitem{Dercks:2018eua}
 D.~Dercks, J.~De Vries, H.~K.~Dreiner and Z.~S.~Wang,
 Phys.\ Rev.\ D {\bf 99} (2019) no.5,  055039
 doi:10.1103/PhysRevD.99.055039
 [arXiv:1810.03617 [hep-ph]].
 
 \bibitem{Pouya}
 M.~Bahraminasr, P.~Bakhti and M.~Rajaee,
 [arXiv:2003.09985 [hep-ph]].
 \bibitem{Kling:2020iar}
 F.~Kling,
 [arXiv:2005.03594 [hep-ph]].

   \bibitem{Lee:1977ua}
   B.~W.~Lee and S.~Weinberg,
   Phys.\ Rev.\ Lett.\  {\bf 39} (1977) 165.
   doi:10.1103/PhysRevLett.39.165
 \bibitem{SLIM}
 C.~Boehm, Y.~Farzan, T.~Hambye, S.~Palomares-Ruiz and S.~Pascoli,
 Phys.\ Rev.\ D {\bf 77} (2008) 043516
 doi:10.1103/PhysRevD.77.043516
 [hep-ph/0612228];
 Y.~Farzan,
 Phys.\ Rev.\ D {\bf 80} (2009) 073009
 doi:10.1103/PhysRevD.80.073009
 [arXiv:0908.3729 [hep-ph]];
 Y.~Farzan and M.~Hashemi,
 JHEP {\bf 1011} (2010) 029
 doi:10.1007/JHEP11(2010)029
 [arXiv:1009.0829 [hep-ph]];
 Y.~Farzan,
 Mod.\ Phys.\ Lett.\ A {\bf 25} (2010) 2111
 doi:10.1142/S0217732310034018
 [arXiv:1009.1234 [hep-ph]].
 
 \bibitem{others}
 S. Dimopoulos, Phys. Lett. B {\bf 84}, 435 (1979); B. Holdom and M. E. Peskin, Nucl. Phys. B {\bf 208}, 397 (1982);
  M. Dine and N. Seiberg, Nucl. Phys. B {\bf 273}, 109 (1986); J. M. Flynn and L. Randall, Nucl. Phys. B {\bf293}, 731 (1987);
  P.~Agrawal and K.~Howe,
  JHEP \textbf{12} (2018), 029
  doi:10.1007/JHEP12(2018)029
  [arXiv:1710.04213 [hep-ph]];
  P.~Agrawal and K.~Howe,
  JHEP \textbf{12} (2018), 035
  doi:10.1007/JHEP12(2018)035
  [arXiv:1712.05803 [hep-ph]];
  T.~Gherghetta, N.~Nagata and M.~Shifman,
  Phys. Rev. D \textbf{93} (2016) no.11, 115010
  doi:10.1103/PhysRevD.93.115010
  [arXiv:1604.01127 [hep-ph]];
  M.~Gaillard, M.~Gavela, R.~Houtz, P.~Quilez and R.~Del Rey,
  Eur. Phys. J. C \textbf{78} (2018) no.11, 972
  doi:10.1140/epjc/s10052-018-6396-6
  [arXiv:1805.06465 [hep-ph]].
 \bibitem{Hook:2019qoh}
 A.~Hook, S.~Kumar, Z.~Liu and R.~Sundrum,
 arXiv:1911.12364 [hep-ph].

 \bibitem{Buckley:2014ana}
 A.~Buckley, J.~Ferrando, S.~Lloyd, K.~Nordström, B.~Page, M.~Rüfenacht, M.~Schönherr and G.~Watt,
 Eur.\ Phys.\ J.\ C {\bf 75} (2015) 132
 doi:10.1140/epjc/s10052-015-3318-8
 [arXiv:1412.7420 [hep-ph]].
 \bibitem{Rojo:2015acz}
 J.~Rojo {\it et al.},
 J.\ Phys.\ G {\bf 42} (2015) 103103
 doi:10.1088/0954-3899/42/10/103103
 [arXiv:1507.00556 [hep-ph]].
 
 
 
 
 
  
 

  
 
\bibitem{Kodama:2007aa}
K.~Kodama \textit{et al.} [DONuT],
Phys. Rev. D \textbf{78} (2008), 052002
doi:10.1103/PhysRevD.78.052002
[arXiv:0711.0728 [hep-ex]].

\bibitem{Kobayashi:2012jb}
T.~Kobayashi, Y.~Komori, K.~Yoshida, K.~Yanagisawa, J.~Nishimura, T.~Yamagami, Y.~Saito, N.~Tateyama, T.~Yuda and R.~Wilkes,
Astrophys. J. \textbf{760} (2012), 146
doi:10.1088/0004-637X/760/2/146
[arXiv:1210.2813 [astro-ph.HE]].

\bibitem{Arrabito:2007rq}
L.~Arrabito, D.~Autiero, C.~Bozza, S.~Buontempo, Y.~Caffari, L.~Consiglio, M.~Cozzi, N.~D'Ambrosio, G.~Lellis, M.~Serio, F.~Capua, D.~Ferdinando, N.~Marco, A.~Ereditato, L.~Esposito, S.~Gagnebin, G.~Giacomelli, M.~Giorgini, G.~Grella, M.~Hauger, M.~Ieva, J.~Csathy, F.~Juget, I.~Kreslo, I.~Laktineh, A.~Longhin, G.~Mandrioli, A.~Marotta, J.~Marteau, P.~Migliozzi, P.~Monacelli, U.~Moser, M.~Muciaccia, A.~Pastore, L.~Patrizii, C.~Pistillo, M.~Pozzato, G.~Romano, G.~Rosa, A.~Russo, N.~Savvinov, A.~Schembri, L.~Lavina, S.~Simone, M.~Sioli, C.~Sirignano, G.~Sirri, P.~Strolin and V.~Tioukov,
JINST \textbf{2} (2007), P02001
doi:10.1088/1748-0221/2/02/P02001
[arXiv:physics/0701192 [physics]].
\bibitem{E137} J. D. Bjorken {\it et al.}, Phys. Rev. D {\textbf{38}} (1988) 3375. 
\bibitem{Steigman:2012nb}
G.~Steigman, B.~Dasgupta and J.~F.~Beacom,
Phys. Rev. D \textbf{86} (2012), 023506
doi:10.1103/PhysRevD.86.023506
[arXiv:1204.3622 [hep-ph]].

   
\end{thebibliography}
\end{document}